\documentclass[12pt]{iopart}

\usepackage{amssymb}
\usepackage{graphicx,color,psfrag}
\newcommand{\sect}[1]{\section{#1}}
\newcommand{\re}[1]{(\ref{eq:#1})}

\def\al{\alpha}

\def\ba{\begin{eqnarray}}
\def\bam{\begin{array}}

\def\be{\begin{equation}}

\def\bi{\bibitem}

\def\di{\partial}

\def\De{\Delta}

\def\ea{\end{eqnarray}} 
\def\ee{\end{equation}}

\def\em{energy-momentum }

\def\fr{\frac}

 \def\ga{\gamma}

\def\ha{\frac{1}{2} }

\def\ka{\kappa}

\def\la {\lambda}

\def\La {\Lambda}

\def\LLL{\left[}

\def\nn{\nonumber}

\def\RR{{\cal R}}

\def\RRR {\right]}

\def\si{\sigma}

\def\sim{\simeq}

\def\td{\tilde}

\def\te{\theta}

\def\ts{\textstyle}

\def\tt{\tilde t}

\def\Te{\Theta}

\def\ve{\varepsilon}

\def\vp{\varpi}

\def\1{{\it one}}

\def\2{{\ts{\ha}}}

\def\3 {\ts{\frac{1}{3}}}
\def\4{\ts{\fr{1}{4}}}
\def\6{\ts{\fr{1}{6}}}

\begin{document}

\title[Pressure in LTB cosmologies]{Pressure in Lema\^{i}tre-Tolman-Bondi solutions and cosmologies}
\author{Donald Lynden-Bell$\dag$ and Ji\v{r}\'{i} Bi\v c\'ak$\ddag$}

\address{$\dag$ Institute of Astronomy, Madingley Road, Cambridge CB3 0HA, United Kingdom,}
\address{$\ddag$ Institute of Theoretical Physics, Faculty of Mathematics and Physics, Charles University, V Hole\v{s}ovi\v{c}k\'{a}ch 2, 180 00 Prague 8, Czech Republic}


\begin{abstract} 
Lema\^{i}tre-Tolman-Bondi (LTB) solutions have traditionally been confined to systems with no pressure in which the gravity is due to massive dust, but the solutions are little changed in form if, as in cosmology, the pressure is uniform in space at each comoving time. This allows the equations of cosmology to be deduced in a manner that more closely resembles classical mechanics. It also gives some inhomogeneous solutions with growing condensations and black holes. We give criteria by which the sizes of different closed models of the universe can be compared and discuss conditions for self-closure  of inhomogeneous cosmologies with a $\La$-term.

\end{abstract} 
\submitto{{\it\CQG} {\bf 33}, 075001 (2016)}

\maketitle

\sect{Introduction}
Oppenheimer and Snyder \cite{1} found the solution for a uniform cold sphere collapsing under gravity to form a black hole. More general solutions that can make black holes are contained in the Lema\^{i}tre-Tolman-Bondi metrics \cite{2,3,4}.

The LTB spherically symmetric, non-static solutions of Einstein's equations have been widely used to investigate the formation of the Cauchy, apparent and event horizons around black holes (see, e.g., \cite{5,6}), the formation of black-hole singularities in regions devoid of matter (e.g. \cite{7}), the appearance of naked singularities (e.g. \cite{8,9}), and the subtle analysis of shell-crossing and shell-focusing singularities (e.g.  \cite{10,11}). LTB models have also played an important role in cosmology as exact inhomogeneous models, supplying the framework to study the formation of structures as clusters of galaxies, galaxies and voids within exact non-linear general relativity \cite{12} and generalising, for spherical symmetry, more astrophysical papers on evolution of structures in the Einstein\,--\,de Sitter universe (see, e.g., \cite{13,14}). Most recently, the LTB models were used to illustrate numerically the occurrence of permanent matter density spikes \cite{15}.

The popularity of the LTB cosmologies increased after the discovery of the accelerated expansion when a possibility appeared that one does not need to evoke the concept of dark energy if we live in an inhomogeneous universe (e.g. \cite{16}). However, it appears that simple LTB models are ruled out as an explanation of dark energy when several observational effects are considered simultaneously (see \cite{17}, p. 411 and references therein).

Above we pointed out some important papers showing the role of the LTB solutions, there are also excellent textbooks and monographs available in which the detailed analysis of the LTB models is given and many other references are listed, see \cite{17}\,--\,\cite{20}.  In most of that  work it is assumed that the pressure is zero. The fluid particles then move along time-like geodesics and the equations for dust become very close to their counterparts in classical mechanics. However, the importance of the pressure of the black body radiation at large redshifts has lessened enthusiasm for the LTB derivation of the cosmological equations themselves because of this lack of pressure.  In his famous work \cite{2} Lema\^{i}tre, in contrast to Tolman \cite{3} and Bondi \cite{4}, did consider the Einstein equations for spherically symmetric, non-stationary, and inhomogeneous fluids with pressure. So, for example, the equation \re{2.10} below relating the time derivative of mass to the radial pressure is contained in \cite{2} (cf. Eq. 3.4). Within the explicit solutions Lema\^{i}tre considers pressure in a quasi-static situation, and in the case of the homogeneous static sphere --- the so-called Eddington problem\footnote{Krasi\'{n}ski \cite{21} explains just how fundamental Lema\^{i}tre's contribution \cite{2} is: the mass for spherically symmetric fluids (now called Misner-Sharp mass) is introduced, an anisotropic pressure is admitted, and an attempt is made to explain formation of structures by  an exact model.}.

 More recently, an interesting generalization of the LTB models to include tangential pressures but without each sphere pressing on the next were given by Gair \cite{22,23}.  The tangential pressure is provided by angular momentum which may differ from shell to shell. In the null limit these models generalize the Vaidya metric \cite{24}. The generalized LTB models including pressure were analyzed by using the ADM formalism in \cite{25,26}. The exterior vacuum (Schwarzschild) spacetime described in generalized Painlev\'{e}-Gullstrand coordinates can be joined to the interior LTB region in a single coordinate system.  However, no dynamical solution with pressure was constructed. S\"{u}ssmann and collaborators \cite{27,28} considered a mixture of matter and radiation, $\rho^{rad}=3p/c^2$. They gave some solutions of interest,  but their form of metric in comoving coordinates  omits the gravitational effects of spatial pressure gradients when compared with the Landau and Lifshitz equations given below. However, the pressure gradients are exactly balanced by the divergence of an appropriately chosen anisotropic pressure. Their work was recently employed \cite{29} to illustrate the results on the existence and stability of shells separating expanding and collapsing regions in the LTB models with anisotropic pressures.

Our aims in this paper are firstly to derive the equations of homogeneous
cosmology  including pressure from the LTB approach. This shows how naturally
Hoyle's continuous creation or inflation fit into cosmology. It  also
demonstrates how closely those equations resemble bound and unbound motion in
spherical  Newtonian dynamics and illustrate the gravitational effect of
internal energy. Secondly we show how such effects change typical LTB
pressureless solutions which we give as examples. Thirdly we consider criteria
for the spatial closure of cosmological models due to their own curvature and
apply them to spherical models.  These are contrasted with conditions for bound
motion.

\section{Field equations and characteristic radii for LTB models with pressure}
\label{sec:2}

Following Landau and Lifshitz Classical Theory of fields p. 364, problem 4, with small changes in notation, we write the metric as
\begin{equation}
ds^2=e^{-2\psi}c^2\,dt^2-e^{2\la}d\chi^2-\left[r(\chi,t)\right]^2\,d\,\hat{{\bf r}}^2,
\end{equation}
where $\chi$ is a comoving coordinate that labels the different spheres whose
areas are $4\pi [r(\chi,\,t)]^2$ at time $t$ and $\hat{{\bf r}}$ is the unit
Cartesian radial vector. Once any particular pole for spherical polar
coordinates is chosen we can express $d{\bf\hat{r}}^2= d\te^2+\sin^2\te
d\phi^2.$ Both $\psi$ and $\la$ are functions of $\chi$ and $t$. We denote
$\di/\di (ct)$ by a dot and $\di/\di \chi$ by a prime. The Einstein equations read
($\kappa=8\pi G/c^4$):
\begin{eqnarray}
\hspace{-4em}-\ka T^1_1&=\ka p=e^{-2\la}\left[(r'/r)^2-2\psi' r'/r\right]-e^{2\psi}(2\ddot{r}/r+\dot{r}^2/r^2+2\dot{\psi}\dot{r}/r)-1/r^2,\label{eq:2.2}\\
\hspace{-4em}-\ka T^0_0&=-\ka\rho c^2=e^{-2\la}(2r''/r+r'^2/r^2-2\la'r'/r)-e^{2\psi}(2\dot{\la}\dot{r}/r+\dot{r}^2/r^2)-1/r^2,\label{eq:2.3}\\
\hspace{-4em}-\ka T^1_0&=0=2e^{-2\la}(-\dot{r}'/r+\dot{\la}r'/r-\psi'\dot{r}/r). \label{eq:2.4}
\end{eqnarray}
As yet the pressure $p\,(\chi,t)$ and the energy density $\rho\,(\chi,t)$ are general.  In particular they include contributions from any cosmological constant so $p=p_m-\La/\ka, c^2\rho=c^2\rho_m+\La/\ka$ where the material has pressure and density $p_m,\rho_m$. The conservation laws come from the contracted Bianchi identities $D_\mu(T^\mu_\nu)=0$ which give
\begin{equation}
2\dot{\la}+4\dot{r}/r=-2\dot{\rho} c^2/(p+\rho c^2)\,,\qquad 
\psi'=p'/(p+\rho c^2)\,. 
\label{eq:2.5}
\end{equation}

The other components of Einstein's equations give nothing new.  The second of equations \re{2.5} shows that if $p$ is independent of $\chi$ then $\psi$ is independent too. Then $e^{-2\psi}$ is a function of $t$ and we may define a new time $\tau$ such that $d\tau=e^{-\psi}c dt$.  The equations are unchanged provided we put $\psi=0$ and reinterpret a dot as $\di/\di \tau$. Equation \re{2.4} reduces to $\di \la/\di\tau =\di(\ln r')/\di\tau$, so on integration $dt$ we find
\begin{equation}
e^{2\la}=r'^{ 2}/(1+2\ve(\chi))\,,
\label{eq:2.6}
\end{equation}
where the denominator is the integration 'constant' which is a function of
$\chi$ alone. $\ve$ turns out to be the energy per unit $mc^2$ of the shell
labelled $\chi$.  Inserting this $e^{2\la}$ into equation \re{2.2} and
remembering that $\psi=0$,
\begin{equation}
-\ka p=-2\ve(\chi)/r^2+(2 \ddot{r}/r+\dot{r}^2/r^2)\,.
\label{eq:2.7}
\end{equation}
But $\di(r\dot{r}^2)/\di\tau=2r\dot{r}\ddot{r}+\dot{r}^3$, so multiplying by $\2\, r^2\dot{r}$, integrating $d\tau$ and defining $M_T,\,M$, we obtain
\begin{eqnarray}
\hspace{-4em}-\ve r+\2 r\dot{r}^2= -\int^\tau\2\ka p(\tau) r^2  \dot{r} d\tau=GM_T(\chi,\tau)/c^2=GM(\chi,\tau)/c^2+\6\La r^3,\label{eq:2.8}\\
\hspace{-4em}M_T(\chi,\tau)=-\int^\tau4\pi r^2 p(\tau)c^{-2}\dot{r}d\tau,~~~M(\chi,\tau)=-\int^\tau 4\pi r^2 p_m(\tau)c^{-2}\dot{r}d\tau,\label{eq:2.9}
\end{eqnarray}
where the integral is performed at constant $\chi$ i.e. over the past history of the shell labelled $\chi$ and includes the integration 'constant' which will depend on $\chi$.  We shall later justify this notation by showing that $M_T$ is the total gravitating mass including the contribution from dark energy. Notice that if $M_T$ is initially zero then {\it all of it is generated by negative pressure} as in Hoyle's  continuous creation and in inflation. Dividing by $r$ and re-ordering the terms, we find an equation with remarkable similarity to the classical energy per unit mass of the shell labelled $\chi$:
\begin{eqnarray}
\2\,\dot{r}^2-GM(\chi,\tau)/(c^2r)-\La r^2/6=\ve(\chi)\,, \label{eq:2.10}\\
\di M_T/\di \tau=-4\pi p(\tau)c^{-2} r^2\dot{r}\,, \qquad \di M/\di \tau=-4\pi p_m(\tau)c^{-2}r^2\dot{r}\nn\,.
\end{eqnarray}
The last two equations are interpreted as the loss (or increase) of mass due to the work done in the expansion of the sphere. We now return to equation \re{2.3} and eliminate $\la$ and its derivatives by using \re{2.6}. The resulting equation involves $\ve(\chi)$ and its derivative but these may be eliminated by use of \re{2.10}.  After a cavalcade of cancellations (see Appendix) we are left with the pleasing result
\begin{equation}
M'=4\pi r^2 \rho_m(\chi,\tau) r',
\label{eq:2.11}
\end{equation}
which justifies our notation and  the interpretation of $M_T$ above. However, it should be realised that this $M$ is not the sum of the $c^{-2}$ times the energy densities within the sphere, because the element of volume is not $4\pi r^2 dr$ (except in the flat case). Indeed $M$ contains a negative contribution from the gravitational binding energy. We are used to the condition $\ve<0$ as being the condition for bound motion but this is no longer true in the presence of the $\La$-term. Rather the condition $\dot{r}=0$ occurs when $\ve=-\left[(GM/c^{2})/r+\La r^2/6\right]$. When there is no pressure, $M$ reduces to the integration 'constant' $M(\chi)$ and then the quantity in square brackets has a minimum at $r=\left(3GM/c^2\La\right)^{1/3} $, so the motion will be bound if $\ve<-(3/2)\left[(GM/c^2)^{2}\La/3\right]^{1/3}$.  For positive material pressure $M$ is not constant (it decreases as the system expands); still this same criterion for bound motion can be used provided $M$ is interpreted as the $M$ at the time when $\dot{r}^2$ has a minimum. Incorporating the value of $e^\la$ found above, the metric now reads
\begin{equation} 
ds^2=d\tau^2-\frac{r'^2d\chi^2}{\left[1+2\ve(\chi)\right]}-r^2\,d\,\hat{{\bf r}}^2.
\label{eq:2.12}
\end{equation}
In considering spatially closed models we show below that there will be a radius at which $r'=0$; and by analogy with the homogeneous cosmological solutions we say that the universe at such points is half-closed. To avoid a metric singularity there, this must occur where $\ve(\chi)=-\2\,$. But since $\ve$ is a function of $\chi$ alone, such a comoving sphere has $\ve=-\2$\, for all time and to avoid metric singularities $r'$ has to remain zero on this comoving sphere. We now ask how this spherical shell with $\ve=-\2\,$ moves. If it emerges from the Big Bang it may  gravitationally decelerate sufficiently for the expansion to cease before the $\La$-term starts accelerating it again\footnote {Surprisingly, larger mass be it due to radiation or rest mass helps rather then hinders escape to infinity. This is because larger masses at fixed $\ve$ emerge from the Big Bang with greater $\dot{r}^2$.}. In such a case its $\dot{r}$ will become zero so from equation \re{2.10} this will occur when $M$ is related to $r$ by $GM/c^2=-\6 \,\La r^3+\2\, r$. We may characterise the extent of such solutions by the maximum radius that the $\ve=-\2\,$ sphere reaches, or alternatively by the gravitating mass $M$associated with that sphere as it reaches that extent. There will also be solutions that collapse from infinity and bounce on the repulsion of the $\La$-term and then expand to infinity. For them there will be a characteristic minimum radius and an associated gravitating mass. However, a third class of spatially closed solutions start from the Big Bang and expand for ever. For them there is no turning point, but they start decelerating after the Big Bang under the influence of gravity but, before they reverse, the effect of the $\La$-term re-accelerates them. There will be a time and a radius at which the $\ve =-\2\, $ sphere has no acceleration. At that moment we see from equation \re{2.7} that $(-\ka p_m+\La)r^2=1+\dot{r}^2$. But combining this with \re{2.10} for $\dot{r}^2$ we find
\begin{equation}
r^3=\frac{3 G M/ c^{2} }{\La- 3\ka p_m /2}\, .
\label{eq:2.13}
\end{equation} 
Here $M(\chi,\tau)$ is to be evaluated for the $\chi$ at which $\ve=-\2\,$ and with $p_m$ at the moment at which that sphere has no acceleration. For a universe like ours $p_m$ was negligible at this time as compared with $\La$ so the above equation effectively relates the characteristic gravitating mass to the characteristic radius.

If the metric describes a self-closed system, then there must be one point (or
an uneven number of points) where $r'=0$. To avoid singularities in the metric
we need $\ve(\chi)=-\2\,$ there. We shall label the (lowest) value of $\chi$ at
which this happens $\chi_1$, so $\ve(\chi_1)=-\2\,$.  It would be possible to
have a closed system in which beyond $\chi_1$ the system is symmetrical with
$r(\chi_1+\De\chi,\tau)=r(\chi_1-\De\chi,\tau)$ with both $\ve(\chi)$ and
$M(\chi,\tau)$ symmetrical about $\chi_1$. Notice that $M'=0$ at $\chi_1$, but
increases up to there and thereafter $M$ starts to decrease as $\chi$ increases
past $\chi_1$.  This decrease of the gravitating mass of material within a
sphere as $\chi$ is increased beyond $\chi_1$ can be interpreted as due to the
increase in binding energy exceeding the increase in rest energy from the
addition of another shell of matter.

\section{The equations of homogeneous cosmology}
We look for solutions of equations \re{2.10} and \re{2.13} in which the radius of any sphere at any time is just a re-scaled  model of the behaviour of any other sphere at that time. Thus we look for solutions of the form $r=a(\tau)f(\chi)$. Notice that replacing $a$ by $a/L$ and $f$ by $Lf$, with $L$ constant, does not change $r$ and furthermore that we are still free to choose our labelling $f$ of the different spherical shells. Inserting this form into \re{2.10} divided by $f^2$, and into \re{2.11}, we get
\begin{eqnarray}
\2\,\dot{a}^2-\6 \, \La a^2=\left[GM(\chi,\tau)/(a f c^2)+\ve(\chi)\right]/f^2,\nn\\
\di M/\di\tau=-4\pi a^2\dot{a}f^3p_m(\tau)/c^2,\, \label{eq:3.14}\\
M'=4\pi a^3\rho_m f^2f'\nn\,.
\end{eqnarray}
From here we deduce that $M=\mu(\tau)f^3$  and $2\ve=-Kf^2$  with $K$ constant and 
\begin{eqnarray}
\mu=\3 \,  4\pi a^3 \rho_m\,, \qquad \dot\mu=-4\pi a^2\dot{a}p_m(\tau)/c^2\,, \label{eq:3.15}\\
\dot{a}^2/a^2-\La/3=8\pi G \rho_m/(3c^2)-K/a^2\,,\label{eq:3.16}
\end{eqnarray}
from which we see that $\rho_m$ has to be a function of $\tau$ only. Cosmologists often use the freedom to choose $L$ described above to rescale $a$ so that for $|K|>0$ the old $|K|/a^2$ becomes the new $1/a^2$. Thus effectively we may take $K=k=\pm 1,0$.  Putting the form for $r$ and $\ve$ into \re{2.7},
\begin{equation}
 \La-\ka p_m=k/a^2+(2\ddot{a}/a+\dot{a}^2/a^2)\,. \label{eq:3.17}
\end{equation}
Equations \re{3.16} and \re{3.17} are the standard cosmological equations and the metric now has $g_{11}=-a^2f'^{ 2}/(1-k f^2)$. For $k=+1$ we may now choose the labelling of our spheres to be $f(\chi)=\sin\chi$, and $f=\sinh\chi$ for $k=-1$, and $f=\chi$ for $k=0.$ With these the metric is
\begin{equation}
 ds^2=d\tau^2-a^2\left[d\chi^2+f^2 \,d\,\hat{{\bf r}}^2\right]. \label{eq:3.18}
\end{equation}

\subsection{Cosmological solutions}
Our solutions contain the cosmological constant explicitly, so our pressure
term is only $p_m(\tau)$, the matter pressure  which is important during the
relativistic/radiation era when $p_m=\3 \rho_mc^2\propto a^{-4}$. We now
concentrate on the special case of a radiation universe with a $\La$ term since
this gives an explicit solution. The more general cosmological solutions give
qualitatively similar results when cold dark matter and baryons are included.
Equation \re{3.16} then takes the form with $\si,\td{\si},\td{k}$ defined below
\begin{eqnarray}
a^2 \dot{a}^2=\3 \, \La a^4-ka^2+\si=\3 \,  \La \left[(a^2-\td{k})^2+\td{k}^2(\td{\si}-1)\right]=F(a^2)\,,\label{eq:3.19}\\
 \td{k}=3k/(2\La)\,,\qquad \si=\3 \,  \ka\rho_mc^2a^4=\mathrm{const}\,, \qquad\td{\si}=4\si\La/(3k^2)\,.\nn
\end{eqnarray}
Notice that when $\td{\si}\ge 1,\,\dot{a}$ is never zero, so the universe either expands for ever or contracts for ever, but when $\td{\si}<1$ reversals will take place at positive $a^2$.  Putting $T=2\sqrt{\3 \, \La}\,\tau$, equation  \re{3.19} integrates via the substitutions $a^2-\td{k}=\td{k}\sqrt{\td{\si}-1}\sinh\chi, \td{\si}>1$ and $a^2-\td{k}=\td{k}\sqrt{1-\td{\si}},\td{\si}<1$ to give
\begin{eqnarray}%
 a^2&=\td{k}\,\left[1\pm \sqrt{\td{\si}-1}\sinh(T-T_0)\right],\qquad & \td{\si} >1,\nn\\
	   &=\td{k}\left[1+ \exp(\pm (T-T_0))\right],                 & \td{\si} =1\,,\label{eq:3.20}\\
	&=\td{k}\,\left[1\pm\sqrt{1-\td{\si}}\cosh(T-T_0)\right],  & \td{\si} <1\,,\nn\\
  	&=\sqrt{3\si/\La}\sinh(T-T_0),                & k        =0\,\nn,\\
  	&=k\,\left[\si-(\tau-\tau_0)^2\right],                     &|k|       >0\,,\La=0\nn\,,\\
  	&=2\sqrt{\si}\,(\tau-\tau_0)\,,                   &|k|       =0\,,\La=0\,.\nn
\end{eqnarray}%
  By suitable choices of $T_0$ it is possible to set $T=0$ when $a=0$ for most, but not all of these solutions. Thus
\begin{eqnarray}%
  a^2&= \td{k}\, \left[\pm\sqrt{\td{\si}}\sinh T-2\sinh^2(T/2)\right] ,\qquad &|\td{\si}-1|>0\,,\nn\\
	 &=\sqrt{3\si/\La}\sinh T, &k=0\,,\nn
\end{eqnarray}%
where only positive solutions for $a^2$ are allowed. 

The nature of these solutions is best seen graphically. Figure 1 shows the graph of $ a^2\dot{a}^2=F(a^2)$ as a function of $a^2$. It is drawn for $\td{k}>0,\,\td{\si}>1,$ so that all positive values of $a$ are accessible with $F>0$.  When $\td{\si}<1$ the only change is that the horizontal axis moves up to the level such as that of the dotted line, then only the two regions drawn with a continuous line are accessible, so there are solutions between $a=0$ and the first $O$-point which then return to the origin and also solutions between the second $O$-point and infinity.  These solutions collapse from infinity but bounce due to cosmic repulsion at that $O$-point and then expand to infinity.  When $\td{k}\le 0$, the minimum of $F$ moves to negative $a^2$, so for real $a,\, F$ increases with $a^2$ and at the origin $F=\3 \,\La\td{k}^2\td{\si}$, which is positive, so all positive values of $a$ are accessible.
\begin{figure}[ht].
   \centering
%
%
\begin{psfrags}%
\psfragscanon%
%
\psfrag{s03}[t][t]{\color[rgb]{0,0,0}\setlength{\tabcolsep}{0pt}\begin{tabular}{c}$a^2$\end{tabular}}%
\psfrag{s04}[b][b]{\color[rgb]{0,0,0}\setlength{\tabcolsep}{0pt}\begin{tabular}{c}$F(a^2)$\end{tabular}}%
%
\psfrag{x01}[t][t]{0}%
\psfrag{x02}[t][t]{1}%
\psfrag{x03}[t][t]{2}%
\psfrag{x04}[t][t]{3}%
\psfrag{x05}[t][t]{4}%
%
\psfrag{v01}[r][r]{0}%
\psfrag{v02}[r][r]{2}%
\psfrag{v03}[r][r]{4}%
\psfrag{v04}[r][r]{6}%
\psfrag{v05}[r][r]{8}%
%
\includegraphics[width=8cm]{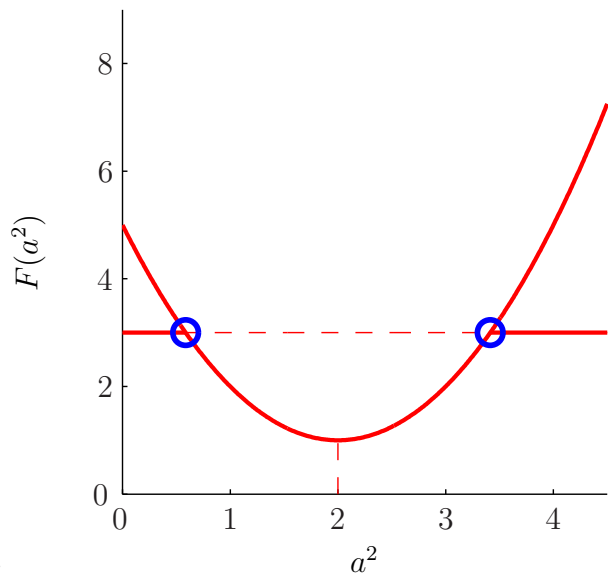}%
\end{psfrags}%
%

\caption{$F(a^2)=a^2\dot{a}^2$ is drawn with the dashed vertical line $\3 \,\La \td{k}^2(\td{\si}-1)=1$ and with $\td{k}=2$. When $\td{\si}<1$ the horizontal axis moves up to a level such as the dashed horizontal line here drawn for $\3 \,\La \td{k}^2(1-\td{\si})=2$, i.e. 3 above its former level.}
\end{figure}
[A very similar graphical analysis holds for the more general case that includes cold dark matter but there the analysis is in terms of $a$ rather than $a^2$ to wit, $\dot{a}^2=\td{F}(a) =\3 \, \La\, a^2+\si a^{-2}+(\3 \,8\pi G\rho_d c^{-2} a^3)/a+k$.]

When there is no $\La$-term the solutions \re{3.20} become
\begin{eqnarray}%
  a^2=k\tau\,\left[2\sqrt{\si}-\tau\right],\qquad\qquad &|k|>0\,,\label{eq:3.21}\\
  a^2=2\sqrt{\si}\tau\,,&k=0\,, \label{eq:3.22}
\end{eqnarray}%
where we have chosen the expanding solutions and set a possible $\tau_0=0$. In the $k=0$ solution the pressure $p_m\propto\tau^{-2}$ and this holds also for the other solutions when $\tau$  is small.

\section{General LTB solutions and global pressure effects}
In this section we give an example of an inhomogeneous pressureless  closed LTB solution that develops a black hole at the origin. We then consider its modification when a $\La$ term is included and finally the effects of a homogeneous pressure. 

 A pretty example is given by setting $x=M/M_U=\sin^3\chi$,  where $M_U$ is the constant gravitating mass of half the closed universe (i.e. that with $\chi\le\pi/2$) and also taking
\begin{eqnarray}%
\ve & =-\2\, \sin^2\!\chi/\,\left(H[x(\chi)]\right)^{2/3},  \qquad \al,\,C,\,x_1=\mathrm{const}>0,  &  \label{eq:4.23}\\ 
H(x)   & =\LLL C+\frac{x^\al}{x_1^\al(1-x^\al)^4+x^\al}\RRR (C+1)^{-1}, \qquad \chi\le\pi/2, \nonumber\\ 
    & = 1, \qquad\qquad\qquad \qquad \qquad\qquad\qquad \pi/2\le\chi\le\pi\,\label{eq:4.24}.
\end{eqnarray}%
 We take $0<\al<1$ and notice that at $\chi=\pi/2,\,x=1,\,H=1$. For general $\La$ and $p_m=0$ we see from equation (10) that the radius of the sphere with gravitational mass $M/M_U=x=\sin^3\chi$ at the 'time' $\tau$ since the Big Bang is given via the integral $\tau=\int dr/\dot{r}=\int^r [2\ve r^2+2G M c^{-2}r+\3 \La r^4]^{-1/2}r dr$,  cf. equation \re{2.10}.

 When $\La=0$ this can be exactly integrated and for $\ve<0$ it gives $r(\tau)$ parametrically in terms of parameter $\eta$:
\begin{eqnarray}%
 r=\left[G M c^{-2}/(-\ve)\right]\sin^2\eta\,,& \label{eq:4.25}\\
 2\pi \tau/\tau_U=\left[2\eta-\sin(2\eta)\right]H(x)\,,\qquad\tau_U=2\pi G M_U c^{-2}\,.\label{eq:4.26}
\end{eqnarray}%
For  $\La=0,\,\tau_U$ is the 'time' between the Big Bang and the Big Crunch but that is not true for  $\La$ not zero.  A singularity forms at the origin where $\eta=\pi$ at time $\tau/\tau_U\ge C/(C+1),  C\ll 1$, and its mass grows almost self-similarly when  $x$ is in the range $C^{1/\al}\ll x/x_1\ll 1$. The constant $x_1$ gives the mass at which the self-similar growth of the singularity ceases. (A suitable special case is given by taking the constant $\al=1/2$.) The universe becomes uniform again as $\chi$ increases towards $\pi/2$, where $M$ approaches $M_U$, and it remains uniform in $\pi/2\le\chi\le\pi$.  The elimination of $\eta$ is easy near the singularity and also near $\eta=\pi/4,\,\pi/2$.  The mass, $M_s$, in the singularity at time $\tau$ is given by setting $\eta=\pi$, so
\begin{equation}
   H(x_s)=\tau/\tau_U\,,\qquad x_s=M_s/M_U\,. \label{eq:4.27}
\end{equation}
Approximating the expression for $H$ above by $\left[C+(x/x_1)^\al\right]/(C+1)$ we solve for the mass in the singularity and find 
\begin{eqnarray}
  M_s=M_1\left[(C+1)\tau/\tau_U)-C\right]^{1/\al},\qquad&\nn\\ 
C/(C+1)\le\tau/\tau_U\ll 1\,, & M_1=x_1 M_U\,;
\label{eq:4.28}
\end{eqnarray}
$M_s$ grows proportionally to $\tau^{1/\al}$ over the self-similar range indicated above. A black hole forms around this singularity. No light can escape from the sphere of radius $r=2GM/c^2$, so we use this rather than the asymptotic definition which for closed universe would give the whole mass of the universe. When the singularity's mass  is small compared with the mass of the universe, the black hole forms where the parameter $\eta$ is near $2\pi$ and in that region $\eta$ is readily eliminated giving
\begin{equation}
    r=\left[G M c^{-2}/(-\ve)\right]\left[(3\pi/2)(1-H^{-1}\tau/\tau_U)\right]^{2/3}.
\label{eq:4.29}
\end{equation}
Setting $2G M c^{-2}/r=1$ we find the mass of the black hole, $M_b$, is given by,
\begin{equation}
   H(x_b)-2x_b/(3\pi)=\tau/\tau_U\,,\qquad x_b=M_b/M_U\,.
\label{eq:4.30}
\end{equation}
Comparison with equation (27) shows that the black hole has a mass only a little more than its central singularity; for example, for $\al=\2$ and $C^2\ll x/x_1\ll 1,\ x_s/x_1\simeq(\tau/\tau_U)^2$ then $M_b/M_s\simeq[1+\frac{4\tau}{3\pi\tau_U}\,x_1]$.

\subsection{Effect of cosmic repulsion}
When the $\La$-term is present the black hole formation is hardly affected. The integral for the time can be well approximated by setting $\La r^4/3=Ar^2+Br$ in the region  of the turn-around radius $r=r_0$ which is the relevant root of the cubic 
\begin{equation}
 2\ve r_0+2GMc^{-2}+\3 \,\La r_0^3=0\,.
\label{eq:4.31}
\end{equation}
  We choose $A= \La r_0^2$ and $B=-2\La r_0^3/3$ so that this term  has the right value and gradient there.  With this approximation the integral becomes
\begin{equation}
  \tau=\int^r\frac{ r dr}{\sqrt{(2\ve+\La r_0^2)r^2+2(G M c^{-2}-\3 \, \La   r_0^{3} ) }}\; .
\label{eq:4.32}
\end{equation}
Comparing this with the $\La=0$ case solved above, $\ve$ is replaced by $\ve+\2\,\La r_0^2$ and $G M c^{-2}$ is replaced by $G M c^{-2} -\La  r_0^3/3$. Thus the $\La$-term decreases the effective binding energy and the effective mass, while $r_0$ itself is approximately $G M c^{-2} (-\ve)^{-1}\left[1-\6 \, \La  G^2 M^2c^{-4}(-\ve)^{-3}\right]$ when the term involving $\La$ is small, but when it is not, one must take the relevant solution to the cubic (see Appendix A2).  Near the singularity the density, $(4\pi r^2)^{-1} dM/dr$, and the contravariant radial velocity component are given by
\begin{eqnarray}
 \rho_m=\frac{x_1^{3/2} c^3}{(2 G)^{3/2}M_U^{1/2}} r^{-3/2}, \label{eq:4.33}\\
 u^r/c=\dot{r}=-\sqrt{2GMc^{-2}/r}\;\,. \label{eq:4.34}
\end{eqnarray}
 The latter is proportional to $r^{-1/2}$ and becomes minus one at the black hole.  When $x\gg x_1$ and when $\chi \ge \pi/2,\,H=1,$ and we find that the density is uniform in space but depends on time, to wit  $8\pi G\rho_m/c^2=3(2G M/c^2)^{-2}\sin^6\eta$ with $\eta$ related to time via \re{4.26} with $H=1$. In these regions the universe expands or contracts uniformly.

 By contrast  the $\La$-term  is really important when $ \La>(3GM_U/c^2)^{-2}$ as there is no turnaround radius for the $\ve=-\2$ sphere so the closed universe expands for ever leaving behind the sphere with $(-2\ve)^{3/2}=3 G M c^{-2}\La^{1/2}$. All material with smaller $M$ eventually falls back  into the black hole.  Using equations \re{4.23} and \re{4.24} with $C$ neglected, the asymptotic mass of the black hole is $M_b=M_1/(3GM_Uc^{-2}\La^{1/2}-1+4x_1^\al)^{1/\al}$.  To get this formula the factor $(1-x_b^\al)^4$ in $H(x_b)$ has been approximated as $1-4x_b^\al$.

\subsection{The gravity of internal energy}
 To get equation \re{2.6} we took the pressure at each cosmic time to be uniform. This is always true of the pressure due to the $\La$-term and true of material pressure in homogeneous cases. In inhomogeneous situations it only occurs astrophysically when radiative cooling is strong so that denser regions cool to give pressure equality. When applied to LTB solutions pressure uniformity removes all forces due to pressure gradients but it leaves the gravity of the changing internal energy and it is this that changes $M$ due to the external work done. Here we briefly consider how such effects influence LTB solutions.  We  turn to the second equation \re{2.10} which is of the same form whether or not there is a pressure $p_m(\tau)$, the only difference being that $M$ depends on $\tau$ as well as $\chi$. Let us consider a sphere  at the moment when it turns around and compare it with a sphere of the same $\ve(\chi)$ and the same $r$ which is also at its turn-around in a pressureless LTB solution. Then the two $M$ must be the same. However, during the subsequent collapse the $M$, in the case with pressure, will increase, provided $p_m$ is positive, since $\dot{M}=4\pi p_m c^{-2} r^2(-\dot{r})$, so the subsequent collapse rate will only be enhanced by the $p(\tau)$ term.  When $\ve\ge 0$ the second equation \re{2.10} together with the requirement that $M\ge 0$ shows that no turning points are possible so everything expands (or in a shrinking universe collapses).  When there is no cosmical constant (or when its effects are negligible), a negative $\ve(\chi)$ ensures that there will be a turning point where $r=-\ve/(GM)$, so such systems will collapse whether or not they are within an ever expanding universe. 

The considerations at the end of Section 3 suggest that the pressure $p_m=\vp\tau^{-2}$ is a natural choice when $\La=0$ and generally when $\tau$ is small.  Then equation \re{2.7} becomes
\begin{equation}
-\ka \vp/\tau^2=-2
\ve Z^{-4/3}+(4/3)\ddot{Z}/Z\,,\qquad Z=r^{3/2}.
\label{eq:4.35}
\end{equation}
We have already shown that $\ve<0$ leads to collapse so we here treat the marginal case $\ve=0$, which is exactly soluble since the equation is linear in $Z$. The ansatz $Z\propto\tau^{s+1/2}$ gives $s^2 -\4 +(3/4)\ka \vp=0$, so $s=\pm \2\, \sqrt{3\ka \vp-1}$ and the general solution is
\begin{eqnarray}
\hspace{-4em}Z&=\tau^{1/2}\left[A(\chi) \tau^s+B(\chi)\tau^{-s}\right] , \qquad r^3=\tau\left[A\tau^s+B\tau^{-s}\right]^2,\quad&3\ka \vp>1\,,\label{eq:4.36}\\
\hspace{-4em}Z&=\tau^{1/2}\left[A(\chi)+B(\chi)\ln(\tau/\tau_0)\right], &3\ka \vp=1\,,\nn\\
\hspace{-4em}Z&=\tau^{1/2}\left[A(\chi)\sin(\td{s}\ln(\tau/\tau_0(\chi))\right],\quad\; \td{s}=\2\,\sqrt{1-3\ka \vp}\;,&3\ka \vp<1\,.\nn
\end{eqnarray}
For these solutions we can find the gravitating masses $M(\chi,\tau)$:
\begin{eqnarray}
\hspace{-4em}M&=(4/3)\pi\vp\tau^{-1}\frac{1+2s}{1-2s}\LLL  A\tau^s-\frac{1-2s}{1+2s}\,B\tau^{-s}\RRR^2+M_1(\chi)\,, \quad &3\ka\vp > 1\,,\nn\\
\hspace{-4em}&=\tau^{-1}\left[A+B\ln(\tau/\tau_0)\right]^2+M_1\,,& 3\ka\vp=1\,,\label{eq:4.37}\\
\hspace{-4em}&=(4/3)\pi \vp A^2 \tau^{-1}\sin^2\left[2\td{s}\ln(\tau/\tau_1(\chi))\right]+M_1\,,&3\ka\vp<1\,,\nn
\end{eqnarray}
where $M_1,\,A,\, B,\, \,\tau_1$ are  'constants' of integration dependent on $\chi$. The unperturbed cosmological solution has $B=0,\,3\ka\vp=1$ and we see that the $B$ solutions blow up relative to that cosmological solution near the Big Bang.

\section{Characterisation and closure with spherical topology}
	To characterise and compare different model universes we need some measure of how big or how massive they are. For a closed pure radiation $\La=0$ universe the total entropy provides a conserved natural measure of its extent. Likewise for a pure dust closed universe the total gravitational mass within the $\ve=-\2\,$ sphere provides a conserved natural measure. However when we ask for a conserved natural measure of a universe containing both dust and radiation, we can not sensibly add entropy to mass and no natural conserved physical quantity replaces them. A way out of this difficulty is to use the maximum radius of the $\ve=-\2\,$ sphere. While that works well when the cosmical constant is zero, many closed universes with $\La$ have no such radius. However the characteristic radius defined by zero acceleration considered in section \ref{sec:2} can be used to take over from the turn-around-radius for those universes that lack the latter and arise from the Big Bang. Furthermore this radius is correctly larger as LTB universes of larger  gravitational mass are considered. Thus although there is no obvious conserved physical quantity associated with this radius the fact that it can be determined from the initial conditions and the equations of motion means that it is a conserved quantity in the sense of dynamics and is thus suitable as a classification parameter.  In the above we considered  the extent of different simple closed spherical universes and showed that they could collapse to a Big Crunch, or to a growing singularity or could have parts that expanded for ever. 

We now consider criteria for a universe to be closed into a spherical topology by its own intrinsic curvature rather than by a  topologist's fiat. A solid angle of $4\pi$ ensures the closure of a two dimensional surface surrounding a point. We have sought but not found an integral over the intrinsic curvature of a 3-surface which gives a criterion for its closure due to that curvature. By analogy with 2-surfaces we expect that an unbounded 3-surface which has a positive lower bound to its curvatures will be closed. Indeed Myers' theorem \cite{30} states that an $n$-dimensional Riemannian manifold is closed if there is no boundary and its Ricci curvature tensor satisfies $R_{\mu\nu}u^\mu u^\nu\ge C_1>0$ for all unit vectors $u^\al$ with $C_1$ constant. However, that  gives only a sufficient condition as there are many surfaces with negative or zero scalar curvatures in places which are nevertheless closed. 
     
For homogeneous, isotropic, pressure-free universes without a $\La$-term all closed models recollapse after expansion; the converse statement is also true\,---\,homogeneous isotropic  models with $p=0=\La$ which recollapse are closed. This, however, is not valid for LTB models. Bonnor \cite{31} considered the LTB cosmologies without a $\La$-term and gave an example of an open universe filled with matter in which every sphere eventually collapses, thus demonstrating that eventual
collapse can occur without spatial closure. So spatial closure  neither implies  eventual collapse nor is it needed for such collapse.

   Now let us turn to the Myers theorem. It was used by Galloway in 1977 \cite{32} to discuss closure for non-rotating, possibly anisotropic, inhomogeneous dust cosmological models.  His theorem can be generalised to apply to the LTB models with spatially uniform pressure and a cosmological constant. Following \cite{32} we take the metric in the form $ds^2=dt^2-\ga_{jk}dx^jdx^k$; in the case of the LTB universes the spatial metric is given by the spatial part of metric \re{2.12}. Introduce a 4-vector $X^\mu$ tangent to the hypersurface $V^3_t (t=$ const) at the point $P$ and then extend it along the flow through $P$ generated by ${\bf u}=\di/\di t$ so that the Lie Bracket $\left[{\bf u,X}\right]=0.$ For the LTB spacetimes with metric \re{2.12} the vector ${\bf X}$ is given by $X^\mu=(0,X^j(r,\te,\phi))$.  Next we calculate the expansion $\Te$ of a small fluid element given by $\Te=u^\mu_{;\mu}$. For the LTB metric with \re{2.12} we obtain $\Te=(\dot{r}'/r')+2(\dot{r}/r)$. The averaged Hubble parameter is defined by $h=\Te/3$. Now let $X=(\ga_{kl} X^k X^l)^{1/2}$ be the length of ${\bf X}$. Then according to \cite{32}, the first two conditions that ensure closure require that at each point $P$ of a section $V^3_t(t=t_0)$ there is recession in all directions, i.e., $\di X/\di t >0$ for all ${\bf X}$, and the rate of recession is decreasing, $\di^2X/\di t^2\le 0$. These conditions are met soon after the Big Bang in the LTB models and the topology of the universe can not change as the models evolve so if the universe is closed, then it remains closed. Expressing $X$ for the LTB models and requiring $\di X/\di t >0$  and $\di^2X/\di t^2\le 0$ for {\it all} ${\bf X}$, we arrive at the following inequalities
\begin{eqnarray}
\dot{r}\ge 0\,,\qquad \dot{r}'r' \ge 0\,,\qquad \ddot{r}<0\,,\qquad r'\ddot{r}'\le 0\,, \label{eq:5.38}\\
r'\ddot{r}'+\dot{r}'+(r'/r)^2(\dot{r}^2+r\ddot{r})-2(r'/r)\dot{r}\dot{r}'\le 0\,.\label{eq:5.39}
\end{eqnarray}
For the homogeneous FLRW universes we just get $\dot{r}\ge 0,\,\ddot{r}<0.$ The conditions considered so far remain unaffected by pressure or by $\La$. The last condition sufficient for closure puts restrictions on the three-dimensional Ricci tensor $\RR_{ik}$ of $V^3_t$ which may be expressed in terms of the spatial components of the four-dimensional Ricci tensor $R_{ik}$ and the time derivatives of $g_{ik}$ giving the extrinsic curvature of $V_t^3$ in the four-dimensional space-time (see, e.g., equations (7), (9) and (10) in \cite{32}).  Now in the presence of a homogeneous pressure that can involve $\La$ we find the spatial components of the four-dimensional Ricci tensor to be given by
\begin{equation}
 R_{ik}=\2\,g_{ik} R+\ka T_{ik}=\2\, \ka (\rho c^2-p)(-g_{ik})\,.
\label{eq:5.40}
\end{equation}
Note that our signature is opposite to that used in \cite{32}, and $-g_{ik}=\gamma_{ik}$ introduced above.  The procedure identical to that described in \cite{32} then implies that the last condition that ensures closure is
\begin{equation}
\mathrm{inf}\,\left[(4/3)\pi G(\rho c^2-p)-h^2\right]=C_1> 0\,,
\label{eq:5.41}
\end{equation}
where $h$ is given by $\Te/3$   and the infimum is taken for all points  in $V_{t_0}^3$.  The LTB model is closed and finite if the conditions \re{5.38}, \re{5.39} and \re{5.41} are satisfied.  Since in \cite{32}  $\La$ is not considered we show what the inequality \re{5.41} implies in the simple FLRW model with dust and $\La$. Then $h^2=(\dot{a}/a)^2$ and the pressure corresponding to $\La$ is $p=-\La/\ka$ and condition \re{5.41} becomes $(\dot{a}/a)^2-\2\, (\3 \, 8\pi\rho/c^2+\La/3)< 0$.  This is only slightly stronger than what follows directly from the exact cosmological equations for FLRW models, namely  $(\dot{a}/a)^2-(\3 \, 8\pi\rho/c^2+\La/3)=-k/a^2$.  Thus the model is closed provided $k=+1.$

 In any closed spherically symmetric universe $r(\chi,\tau)$ must return to zero at $\chi=\pi$ say, so there must be a point where $r'=0$ and we found in section 2 that such a point must be comoving with $M$ reaching its maximum there as a function of $\chi$. When there is no pressure $M$ is a function of $\chi$ alone so this is then an overall maximum.

\section {Conclusions}

    We have explored the derivation of the cosmological models through the Lema\^{i}tre-Tolman-Bondi approach generalized to include homogeneous pressure
term and a non-vanishing cosmological constant. The presence of pressure implies the changing of mass of each individual spherical shell in the LTB
models. If the mass is initially zero then all of it can be generated by negative pressure as in Hoyle's continuous  creation or inflation. The change of the mass of a sphere can be interpreted as the work done in the expansion/contraction of the sphere. We considered especially closed models and pointed out that there are different types of motion depending on the repulsive role of the cosmological term. Using the LTB methods we also obtained explicit homogeneous cosmological solutions with both the cosmological term and pressure due to radiation.
    
In a more general setting we analyzed the pressureless closed LTB solution in which a black hole develops at the origin and then considered the effects of a $\La$-term and of a homogeneous pressure. The $\La$-term  does not affect the black hole formation but affects its final growth particularly when $\La$ is larger than $(3GM_U/c^2)^{-2}$, where $M_U$ is the constant gravitating mass of half the closed universe. Then the universe expands for ever leaving behind the black hole. We give the formula for its asymptotic mass.   
     
Although we assume that the material pressure is time-dependent but spatially homogeneous the gravity of internal energy is changed due to the external work and such effects influence the behavior of the LTB solutions. For example, the collapse rate increases due to the pressure term because mass increases with (positive) pressure.
      
By analogy to the radiation-filled Einstein-de Sitter universe we considered the LTB models with $\La=0$ and pressure $p\propto\tau^{-2}$. Under these assumptions we found the LTB solutions in which the gravitating masses can be explicitly determined. As compared with homogeneous case some of these inhomogeneous solutions blow up near the Big Bang relative to the standard models but others can emerge acceptably from the Big Bang. In more general situations in which the  pressure is close to homogeneous, one can turn to the perturbation theory of the LTB models with pressure
we present.       
      
Finally,  we applied Myers' theorem, well-known in geometry but not used in the LTB context, to give criteria for the self-closure of inhomogeneous spherical cosmologies with a cosmological constant.

\ack{J.B. acknowledges the support from the Czech Science Foundation, Grant No. 14-10625S and the kind hospitality of the Institute of Astronomy, University of Cambridge. We thank David Kofro\v{n} for the help with the paper.}

\appendix
\section*{Appendix}
\setcounter{section}{1}
\subsection{Cancellation cavalcade}
We use equation \re{2.6} to eliminate $\la$ from equation \re{2.3}; this gives
\begin{eqnarray}
\hspace{-4em}\ka\rho_m c^2+\La&=-\frac{(2\ve+1)}{r'^2}\LLL2\left(\frac{r'}{r}\right)' +3\left(\frac{r'}{r}\right)^2-2\,\frac{r'}{r}\left(\frac{r''}{r'}-\frac{2\ve'}{1+2\ve}\right)\RRR + \nn\\
\hspace{-4em}&\quad +\LLL2\,\frac{\dot{r}\dot{r}'}{rr'}+\frac{\dot{r}^2}{r^2}\RRR+\frac{1}{r^2}\,.
\label{eq:6.42}
\end{eqnarray}
We now use equation \re{2.10} to eliminate $\ve$ and obtain
\begin{eqnarray}
\hspace{-4em}-\ka\rho_m c^2-\La/3=\LLL\dot{r}^2-\frac{2GM}{c^2r}-\frac{\La}{3}\, r^2+1\RRR\LLL\frac{2(r'/r)'}{r'^2}+\frac{3}{r^2}-\frac{2r''}{rr'^2}\RRR\ + \nn\\
\hspace{-4em}\quad\frac{2}{r r'}\LLL-\frac{GM'}{c^2r}+\frac{GMr'}{c^2r^2}\RRR-\frac{\dot{r}^2+1}{r^2}\,. 
\label{eq:6.43}
\end{eqnarray}
Now $(2/r'^2)(r'/r)'=2r''/(r r'^2)-2/r^2$ so the second square bracket reduces to $1/r^2$. As a result almost all the terms cancel and we are left with
\begin{equation}
\ka \rho_m c^2=2G M' c^{-2}/(r^2 r')\,,
\label{eq:6.44}
\end{equation}
which gives  $M'=4\pi r^2 \rho_m r' $ as recorded in equation \re{2.11}.

\subsection{Cubic solution}
Exact solution of the cubic for the turn-around radius.  The cubic is
\begin{equation}
r_0^3-3y r_0+2b=0\,;  \qquad y=-2\ve/\La\,, \qquad b=3 G M c^{-2}/\La\,.
\label{eq:6.45}
\end{equation} 
With $\ve<0,\ y>0$, there is one (unphysical) negative root $r_0=r_3$ which is readily found by the standard procedure of writing $r_0= w+y/w$ and solving the tri-quadratic for $w$ that results.  We thus find $r_3= -  b^{1/3}\left[(1+\sqrt{1-Y^2})^{1/3}+(1-\sqrt{1-Y^2})^{1/3}\right],\ Y=y^{3/2}/b$; since this is a root, we can subtract the expression on the left of \re{6.45} evaluated at $r_3$ and divide the result by $r_0-r_3$ to find the quadratic for the other two roots, which are $r_0=\left[-\2\,r_3\pm\sqrt{3(y-r_3^2/4)}\right]$. The smaller of these roots is the turning point we seek; the other root is the minimum radius of the universe that contracts from infinity but bounces on the $\La$-term before returning to infinity.

\end{document}